# Optimization of honeycomb battery package based on space mapping algorithm

Wenquan Shuai[1], Xin Luo[1], Hu Wang[*,1,2],

[1]State Key Laboratory of Advanced Design and Manufacturing for Vehicle Body, Hunan University, Changsha, China

[2]Joint Center for Intelligent New Energy Vehicle, Shanghai, China

Jian Wang[3]

## ABSTRACT

A new honeycomb battery package structure is designed and optimized in this study. It's a honeycomb structure which uses grid to reinforce the strength. To obtain the highly accurate finite element (FE) model, the material parameters of 18650 cylindrical Li-ion battery are identified by using optimization techniques based on flat compression test data. Due to the expensive cost of finite element evaluation, the space mapping (SM) algorithm is suggested to optimize the structure of the package. Compared with other space mapping algorithms, the coarse model of space mapping in this work is based on a pseudo-plane-strain model. Moreover, to guarantee the reliability, the mean and variance values of battery stress are used to be the objective function. The final optimum solution is obtained in 3 days, and it shows the magnitude of stress and the distribution of stress are improved significantly compared with initial structure. Moreover, the computational cost of optimization for the problem is decreased greatly.

[*]Email: wanghu@hnu.edu.cn, Tel 86-731-88821417, Fax 86-731-88821807

**Keyword:** Li-ion battery package; Optimization; Simulation; Space mapping(SM) algorithm

## 1. Introduction

The Li-ion battery is considered to be an ideal electric vehicle battery due to its higher energy density, higher power density and longer cycle life. The safety of the cell should be more valued in its application. When the cell is abused, it might explode or get fire due to internal short circuit [2]. These abuses include mechanical abuse, overcharge, over-discharge and overheating. In this work, only mechanical abuse is concerned. In order to understand the mechanism of internal short circuit, experiments are conducted. However, these experiments are ruinous and unrepeatable. Therefore, FE method is commonly used to simulate and predict the behavior of battery. However, it is difficult to develop the constitutive model of Li-ion battery when building the FE models. In order to solve the problem, a lot of researches had been done. For examples, Sahraei *et al.*, (2012a) performed mechanical test on pouch battery under five loading conditions and used compressible foam to build FE models of pouch battery. Sahraei *et al.*, (2012b) modeled 18650Li-ion battery and obtained the short circuit detection criterion under mechanical abuse condition. They also used the compressible foam to build FE models of cylinder battery but the material parameters were obtained from the strain-stress curve of pouch battery. Wierzbicki and Sahraei (2013) constructed FE models of 18650 Li-ion battery and studied the homogenized mechanical properties of the battery. They conducted the flat compression test and used the principle of virtual work to obtain the strain-stress curve of cylinder battery. Greve and Fehrenbach (2012) performed quasi-static mechanical test on cylindrical battery and built a macro-mechanical FE model to simulate the deformation and short circuit initiation of cylindrical battery. Ali *et al.*, (2013) used computational models of Li-ion battery to simulate the behavior of battery under constrained compression tests. Sahraei *et al.*, (2014) tested three types of pouch battery ranging from small cells to large cells under several different global and local compression conditions to obtain their mechanical property. They also constructed models for these three types of battery and investigated conditions leading to internal short circuit in the cells. Sahraei *et al.*, (2016) developed a micro model of representative volume element (RVE) for Li-ion battery to study the failure mechanisms of internal component under complex loading scenarios. The available model for applying can be created based on these achievements. In this work, the crushable foam constitutive model is used to model the FE model of 18650 battery and the material parameters are obtained by inverse identification method.

In this study, the new battery pack structure is suggested. Compared with the typical honeycomb structure, the distinctive characteristic of this structure is adopting a grid to reinforce the strength [1]. It is well known that typical honeycomb structure is an excellent function structure which has outstanding crashworthiness and energy-absorption property. Therefore it has been wildly used and investigated. Yamashita and Gotoh (2005) utilized numerical simulation to analyze the effect of honeycomb cell shape and foil thickness on crush strength of honeycomb which is compressed in the longitudinal direction. Ivañez *et al.*, (2017) studied the influence of structure size and material properties for energy-absorption property of honeycomb core. Zhang *et al,*. (2017) investigated the indentation response and energy absorption of honeycomb sandwich panels under drop-weight impact behavior and obtained some semi-empirical formulations under different impact energy. However, the typical honeycomb structure is too soft to be used as battery package structure directly. Therefore, its strength should be reinforced by using grid, and the largest difference from typical honeycomb is that the honeycomb is not applied as buffer structure of battery package. The batteries are placed into its hexagon cells, so that the special hexagon cell is utilized to protect internal battery.

In order to obtain better performance of battery package structure, optimization should be conducted. However, the computational cost of FE model is too high. It results in that typical direct optimization algorithm cannot be used in this problem. Space mapping (SM) algorithm is an efficient optimization method, especially for the problems of high calculating cost. The SM algorithm establishes two types of model which are fine model and coarse model, respectively. Fine model is a high accurate model or evaluation, which can achieves a more accurate result but requires highly computational cost. Coarse model is much simpler and achieves a result efficiently, but the accuracy of the result is lower. The SM would establish a mapping relation between response of fine model and coarse model. The method can get the optimized result of coarse model in a short time, and then obtain the optimized result of fine model through the mapping relation. Leary *et al*., (2001) used constraint mapping method to optimize an expensive model. The method mapped the coarse model constraints so that the coarse model constraints could approximate the fine model constraints. It is accurate to apply to an expensive model whose real objective varies little but constraints change signally. Redhe *et al.*, (2002) used the SM algorithm which uses surrogate models and response surface method (RSM) to optimize a series of vehicle crashworthiness problems. Florentie *et al.*, (2016) applied the SM algorithm in fluid-structure interaction problems. The output SM was used to optimize partitioned fluid-structure interaction problems, and resulted in the computational cost decreased to 50% in comparison with quasi-Newton method. Wang *et al.*, (2017) integrated reanalysis method with SM method to propose an available optimization method. The Reanalysis-based space mapping method shows a significant improvement in the efficiency of expensive simulation-based problems. Due to the mentioned advantages above, the SM algorithm seems to be suitable for the problems with highly computational cost. Therefore, the SM is applied to structure optimization of honeycomb package in this study.

The rest of paper is organized as follows. In Section 2, the material parameters of Li-ion battery are obtained by the inverse identification. What's more, the material parameters of Al alloy are obtained from uniaxial tensile test. In Section 3, the modeling method of battery package is validated by experiments and then the FE model of battery package is established. Subsequently, the trust region SM (TRSM) algorithm is introduced and used to optimize the battery package structure. In the final Section, the conclusions are made.

# 2. Identification of material parameters of Li-ion battery and Al alloy 6061-T6

## 2.1 Inverse identification of material parameters of battery

### 2.1.1 Inverse identification method

For the battery package design, it is important to use highly accurate FE evaluation in the optimization procedure. The material parameters are the critical issue for simulation accuracy. Therefore, the hybrid numerical method is used to obtain the material parameters in this study. The flow chart of the inverse identification method is shown in Fig. 1.

### 2.1.2 Flat compression test of 18650 battery

The Li-ion batteries studied in this work are commercial 18650 ternary polymer lithium batteries. Their capacity is smaller than pouch battery's, only 2600mAh. It means that an electric vehicle needs more batteries. However, they are safer than pouch battery. The specifications of these 18650 batteries are listed in Table 1.

The contribution of shell casing to total force is less than 1% in the flat compression test [5]. Therefore, whether there is shell casing or not is not important for the test. In order to measure the voltage of battery, the shell casing is retained, as shown in Fig. 2(a). The compression test is performed in INSTRON 150kN Universal Test Machine, and digital image correlation (DIC) is used to measure the local strain (Fig. 2(c)). The UNI-T U322 Thermometer and UNI-T UT61C Multimeter are used to obtain temperature and voltage respectively.

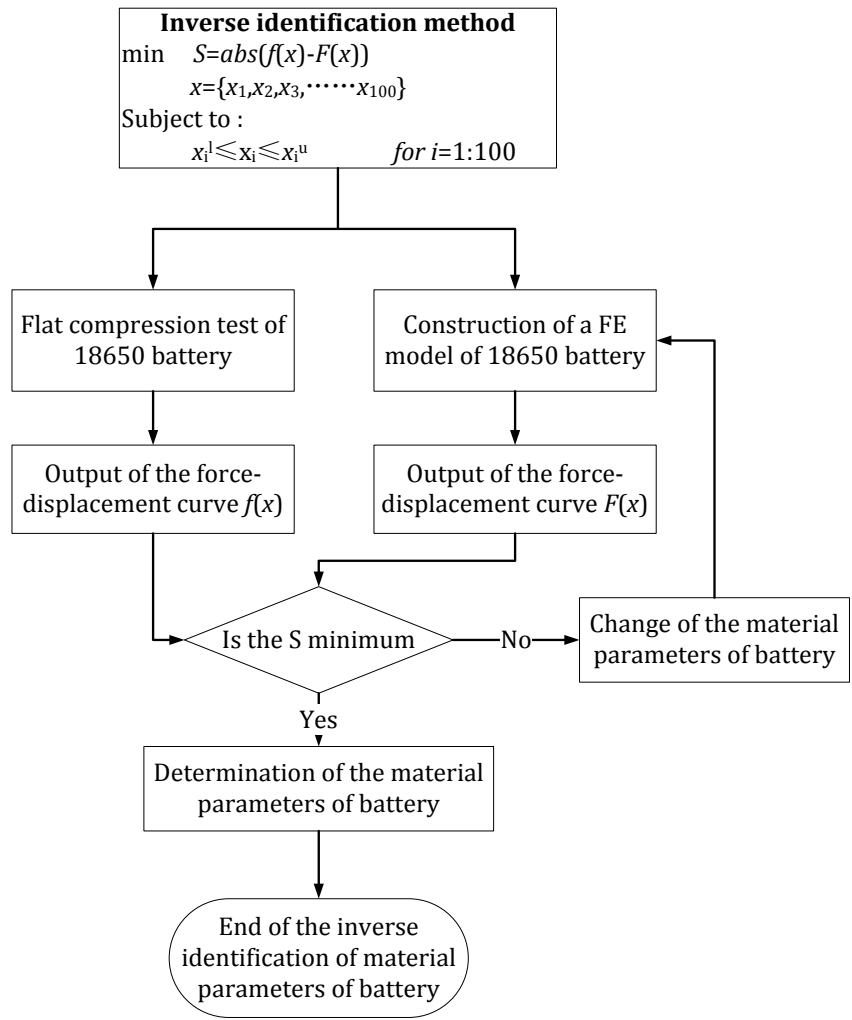

Fig. 1 The flow chart of the inverse identification method

The deformation rate is set as 1mm/min. When the internal short circuit happens, the test should be stopped. The response of internal short circuit is that the load and the voltage drops rapidly and the temperature increases simultaneously [2].

The force, displacement, temperature and voltage are measured in the test. After accomplishment of the test, it is observed that the electrolyte of battery is leaked and the shell casing is flawed. The deformed battery is shown in Fig. 2(d). Figure. 3 shows the data which is obtained in the test. It reveals the load drops rapidly from $5.7 \times 10^4$N when the displacement is about 6.3mm. Meantime, the temperature start to increase significantly and the voltage drops from 3.6V to 0.096V in a short time. It means that internal short circuit has happened.

Table 1 Specifications of the 18650 battery

| Parameters | Values |
|---|---|
| Nominal capacity | 2600mAh |
| Size | 18mm*18mm*65mm |
| Weight | 48g |
| Nominal voltage | 3.7V |
| material | Li(NiCoMn)$O_2$ |
| Operating temperature | -20℃ − 60℃ |

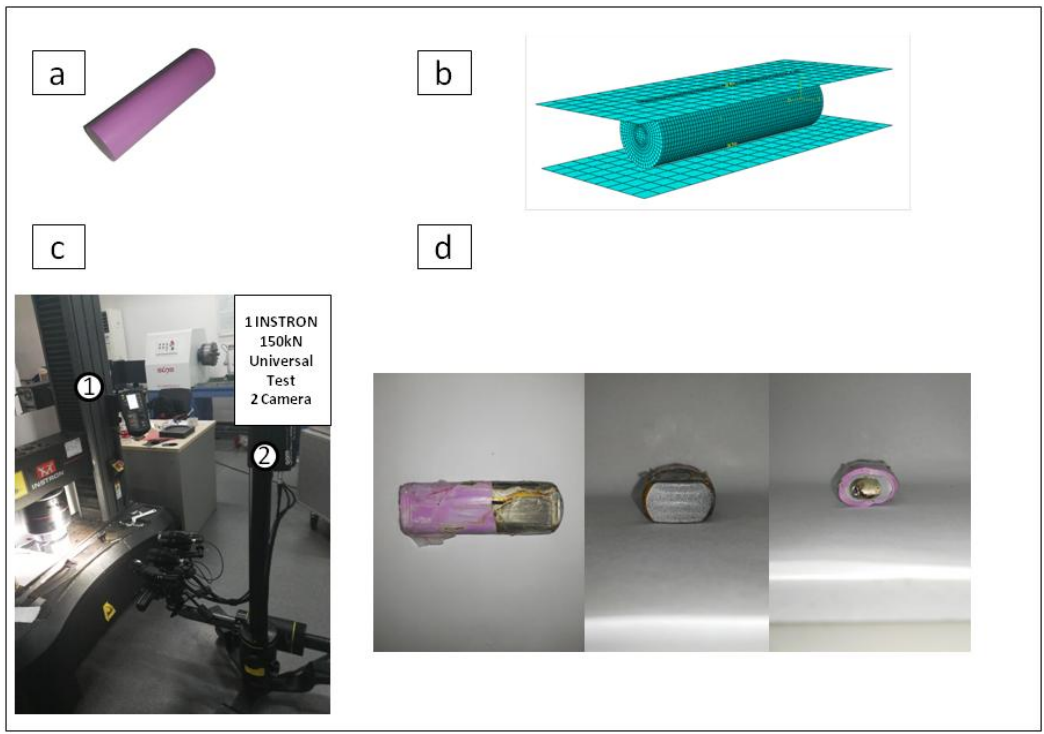

Fig. 2 (a) The entity of battery (b) The FE model of battery (c) The experimental facilities of the flat compression test (d) The deformed battery

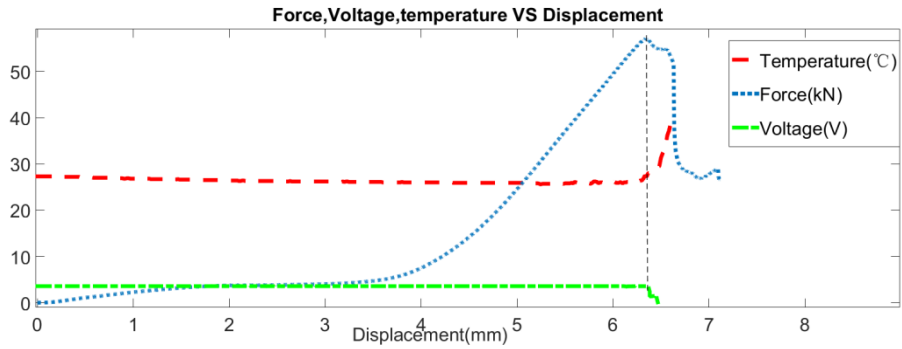

Fig. 3 The result of the flat compression test

### 2.1.3 FE model of 18650 battery

A FE model of 18650 battery is constructed in ABAQUS. The contribution of shell casing to total force is too small, therefore, the whole cell is regarded as a homogeneous and isotropous material and the shell case is ignored. The crushable foam is used as the constitutive model of the material in the plastic phase[2]. To simulate the real compression test, the model of the cell is compressed by two rigid shells. The cell is modeled by using 8-node linear brick and reduced integration solid elements (C3D8R). The simulation parameters are detailed in Table 2. The whole model is shown in Fig. 2(b). The contact force between rigid board and battery is exported as history output.

Table 2 The simulation parameters of battery FE analysis

| Parameters | Values |
| --- | --- |
| Global seed size of the cell | 1mm |
| Global seed size of the rigid shell | 5mm |
| Simulation step | Dynamic explicit; 0.12s |
| Loading speed | 50mm/s |
| Friction coefficient | 0.1 |

### 2.1.4 Inverse identification of material parameters of battery

The material parameters of the cell are obtained by inverse identification method. The two force-displacement curves are obtained by experiments and simulation, respectively. The objective function is minimizing the error between two force-displacement curves, and there are 11 design variables in this problem. The mathematical expression can be formatted as follows

$$\begin{cases} \min_D \sum_{i=1,j}^{100}(f(x_i) - F(x_j))^2 \\ D = [E \ \nu \ \sigma_1 \ \sigma_2 \ \sigma_3 \ \sigma_4 \ \sigma_5 \ \sigma_6 \ \sigma_7 \ \sigma_8 \ \sigma_9] \end{cases} \quad (1)$$

where $x_j$ is the nearest displacement from the $x_i$ in the experimental displacement data. $D$ represents the design variables where $E$ is elasticity modulus, $\nu$ is Poisson ratio, $\sigma_{1\to9}$ is yield stress in harden curve which shown in Table 3. $f(x_1)$ represents the contact force when the displacement is $x_1$ in simulations. $F(x_2)$ represents the force when the displacement is $x_2$ in experiments.

Each FE evaluation outputs 200 sets of data of contact force and time, and the interval between two sources is about 0.03mm. Similarly, the experiment outputs 4312 sets of data of force and displacement. The interval between two sources is nearly 0.002mm. The 100 sets of simulation data are selected to calculate the error. To obtain reliable result, the global evolutionary optimization method, Genetic algorithm (GA) is employed. The final result is listed in Table 4 and corresponding curves are demonstrated in Fig. 4. It can be found that the curves derived from experimental data and simulation are well matched. $R^2$ is a criterion which is applied in fitting nonlinear curves, as shown in Eq. (2) where $y$ represents the true value and $y^*$ represents the predicted value. The criterion $R^2$ is 0.9 in this work. It means the inversed parameters are well matched with experimental data and can be used for subsequent simulation.

$$R^2 = 1 - \frac{\sum_{i=1}^{n}(y_i - y_i^*)^2}{\sum_{i=1}^{n}(y_i)^2} \tag{2}$$

Table 3 The parameters of harden curve of material of battery

| Yield stress/MPa | Plastic strain |
|---|---|
| $\sigma_0$=7.6958 | 0 |
| $\sigma_1$ | 0.010 |
| $\sigma_2$ | 0.025 |
| $\sigma_3$ | 0.065 |
| $\sigma_4$ | 0.100 |
| $\sigma_5$ | 0.125 |
| $\sigma_6$ | 0.165 |
| $\sigma_7$ | 0.215 |
| $\sigma_8$ | 0.255 |
| $\sigma_9$ | 0.325 |

Table 4 The material parameters of battery

| Parameters | Values | Unit |
|---|---|---|
| $E$ | 544.8866 | MPa |
| $v$ | 0.0234 | |
| $\sigma_1$ | 9.4787 | MPa |
| $\sigma_2$ | 7.7454 | MPa |
| $\sigma_3$ | 7.8069 | MPa |
| $\sigma_4$ | 7.7137 | MPa |
| $\sigma_5$ | 7.6989 | MPa |
| $\sigma_6$ | 7.7091 | MPa |
| $\sigma_7$ | 10.1857 | MPa |
| $\sigma_8$ | 106.6688 | MPa |
| $\sigma_9$ | 161.4426 | MPa |

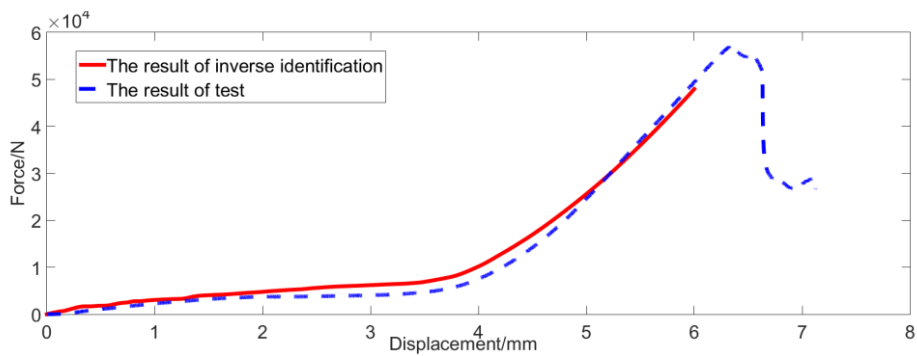

Fig. 4 The force-displacement curves of experiments and simulation

## 2.2 The material parameters of Al alloy 6061-T6

### 2.2.1 Uniaxial tensile test of Al alloy 6061-T6

The material of the wall of battery package is Al alloy 6061-T6. In order to obtain the material parameters of Al alloy 6061-T6, the uniaxial tensile test is conducted. The size of tensile test coupon is obtained according to GB/T 228-2010, as shown in Fig. 5(a). The uniaxial tensile test is conducted in INSTRON 150kN Universal Test Machine. To obtain the strain of tensile test coupon during the test, the extensometer is used. The tensile rate is 2mm/min. Figure. 5(b) shows the fractured tensile test coupon. The engineering strain- stress curve is obtained from test. It should be transformed to true stress and strain. The transformation formula is

$$\begin{cases} \epsilon_T = \ln(1 + \epsilon) \\ \sigma_T = \sigma(1 + \epsilon) \end{cases} \tag{2}$$

The true strain- stress curve is shown in Fig. 6(a).

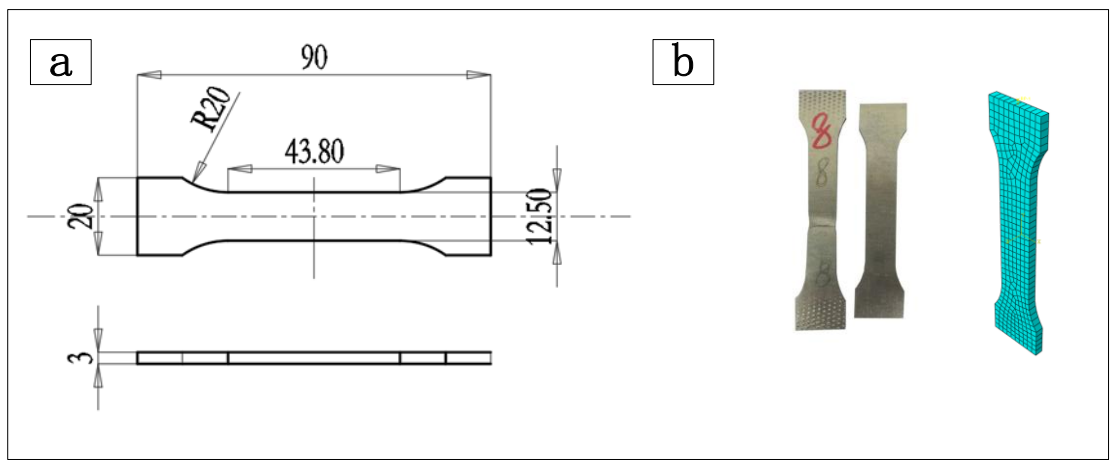

Fig. 5 (a) The size of tensile test coupon (b) The fractured tensile test coupon and corresponding FE model

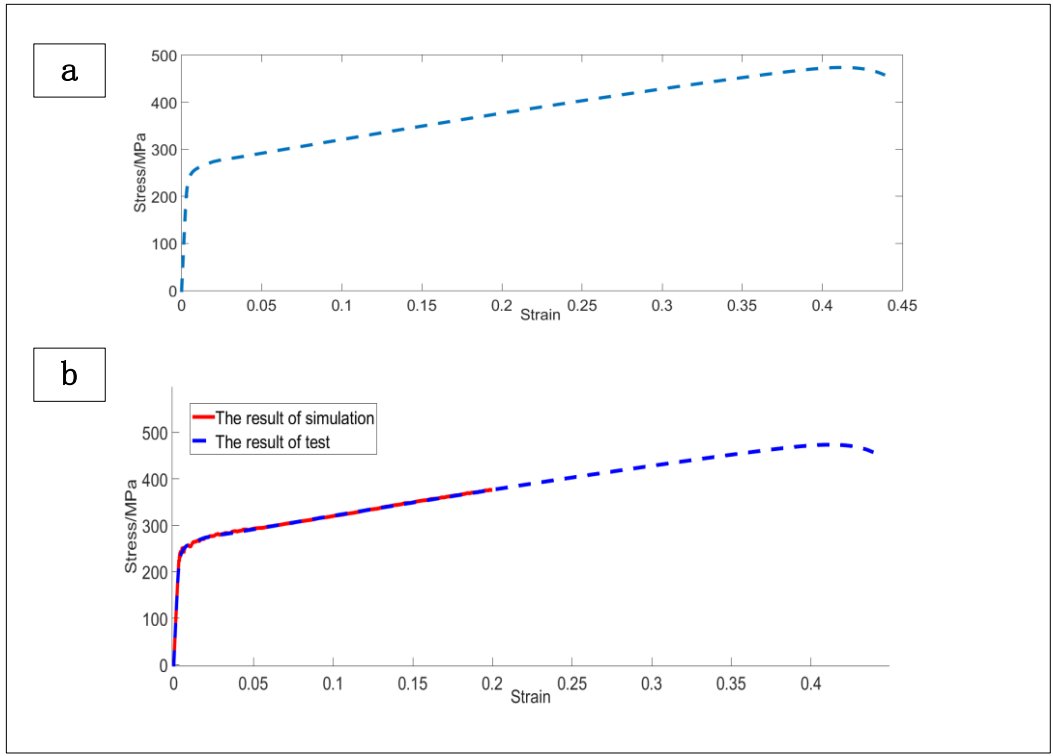

Fig. 6 (a) The true strain- stress curve of tensile test (b) The result curves of parameters validation

## 2.2.2 The validation of material parameters of Al alloy 6061-T6

According to the above illustration, the material parameters can be obtained from the true strain- stress curve. The material parameters of Al alloy 6061-T6 are summarized in Table 5. In order to validate the material parameters, the FE model of tensile sample is built, as shown in Fig. 5(b). The strain-stress curve of elements of FE model is output to validate the accuracy of parameters. The validation result is shown in Fig. 6(b). The result indicates the material parameters of Al alloy are accurate.

Table 5 The material parameters of Al alloy

| Parameters | value | |
|---|---|---|
| Elasticity modulus /MPa | 71.275 | |
| Poisson ratio | 0.33 | |
| Yield strength/ MPa | 241.5 | |
| Density/ (T/mm³) | 2.9e$^{-9}$ | |
| | Yield stress/MPa | Plastic strain |
| | 241.5 | 0 |
| | 263.0 | 0.0069 |
| | 278.8 | 0.0217 |
| | 318.8 | 0.0921 |
| Plastic curve | 346.7 | 0.1408 |
| | 374.5 | 0.1914 |
| | 388.8 | 0.2181 |
| | 423.8 | 0.2862 |
| | 464.3 | 0.3728 |
| | 473.6 | 0.4078 |

# 3. Design and optimization of battery package

## 3.1 The validation of modeling method of battery package

### 3.1.1 Quasi-static compression test of battery package

The computational cost would be expensive if the whole package is modeled by solid elements. We hope to use a specific simplified FE model to complete the package design. In order to validate the modeling method of battery package, a quasi-static compression test is conducted to compare with FE model. The structure of the test battery package without battery is shown in Fig. 7(a). Figure. 7(b) shows the test package sample. The quasi-static compression test is conducted in INSTRON 2000kN Universal Test Machine. The compression rate is 1mm/min. The deformed sample is shown in Fig. 7(d).

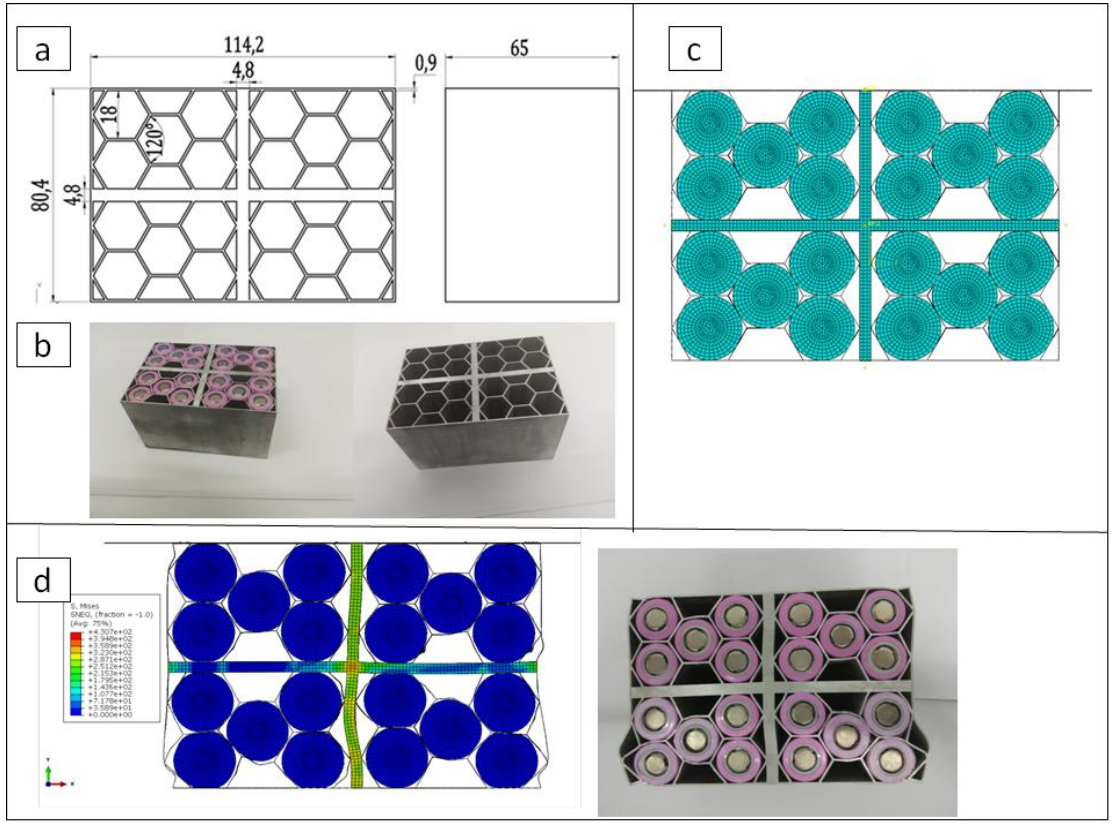

Fig. 7 (a) The structure of the battery package (b) The test package sample (c) The FE model of test sample (d) The stress contours of FE model and the deformed sample

### 3.1.2 The validation of FE model

In order to improve the efficiency of simulation, the FE model is simplified and different from the entity. The FE model is divided into two parts. One is the honeycomb core which is modeled by 4-nodes shell element (S4RSW). Another is the reinforce grid which is modeled by solid element (C3D8R). These two parts are tied together by using tie constraint. The whole model includes 428805 solid and 96980 shell elements. The friction coefficient of model is 0.17. The test is quasi-static compression test. In order to improve its computational efficiency, the compression rate is set as 80mm/s. It can be observed that the kinetic energy is less than 5% of the total energy, therefore, it is still a quasi-static process.

Figure. 7(c) shows the FE model of test sample. The compression displacement is 3mm. The stress contours of FE model is shown in Fig. 7(d). By comparing the two figures in Fig. 7(d), it can be observed that the deformed shape of test and simulation are similar. The bottom honeycomb cores of test and simulation are both buckling outward and the grid both bend. Therefore, the modeling method can be accepted in this work.

## 3.2 The FE model of battery package

The structure of optimized battery package is shown in Fig. 6(a). The FE model of package also consisted of two parts which are honeycomb core and grid. The honeycomb core is modeled by using 4-node shell elements (S4RSW), its thickness is 0.9mm. The grid is modeled by using solid elements (C3D8R). The whole model includes 672165 solid and 135200 shell elements. The friction coefficient of model is 0.17. Figure. 8(b) shows the FE model of the whole battery package. Figure. 9(a) shows the stress contours of the package in 0.018s, 0.036s, 0.054s and 0.072s, respectively. These contours show the stress of model increases with the time increases. In order to show the integral stress contours of battery only, the grid and honeycomb are excluded from the simulation output and only the battery is output as shown in Fig. 9(b). The maximum stress of battery is 32.1733MPa which is still in safe loading range.

The FE model of battery package can simulate the stress-strain behaviors of the battery package accurately, however, the computational cost is high. It needs 2.5 hours to finish a simulation when using 8 CPUs (Inter(R) Xeon(R) CPU,3.40GHz). Therefore, the typical optimization method cannot be available for optimizing the battery package. However, the SM algorithm can do well in this kind of expensive problems. In later section, the TRSM algorithm is introduced and applied in optimization of the battery package.

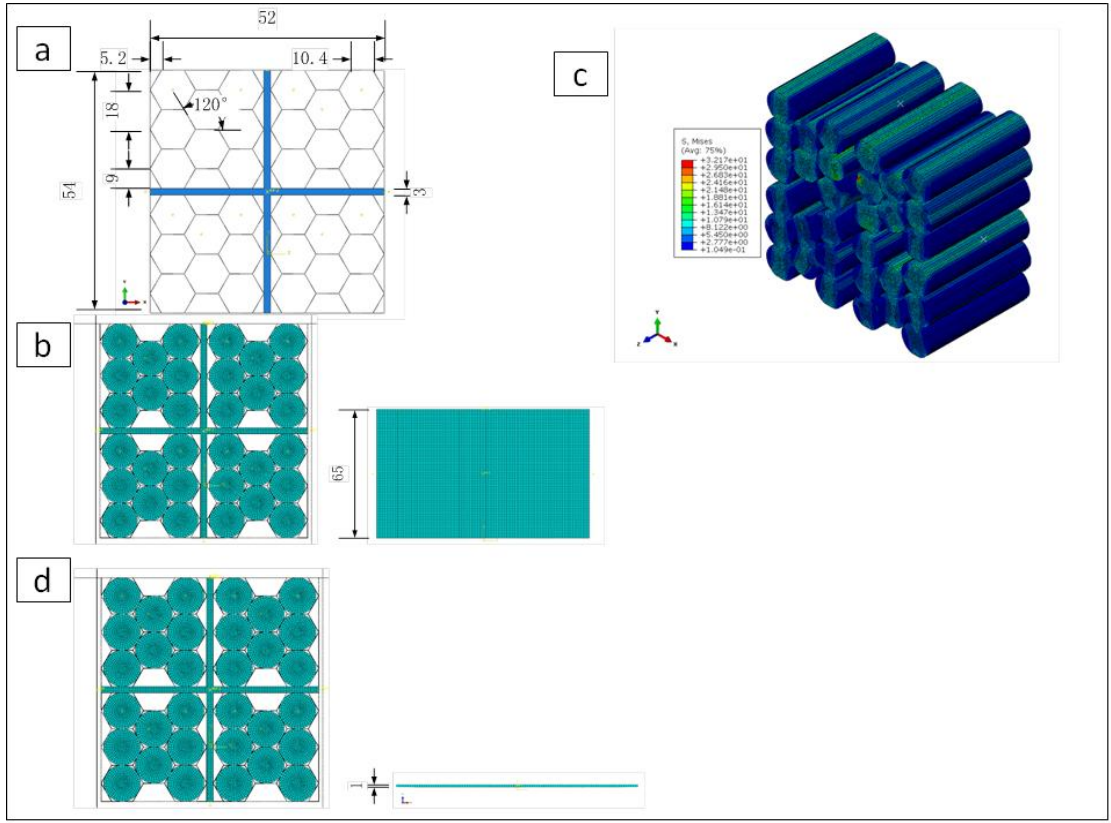

Fig. 8 (a) The structure of optimized battery package (b) The FE model of the whole battery package (c) The stress contours of the batteries (d) The coarse FE model

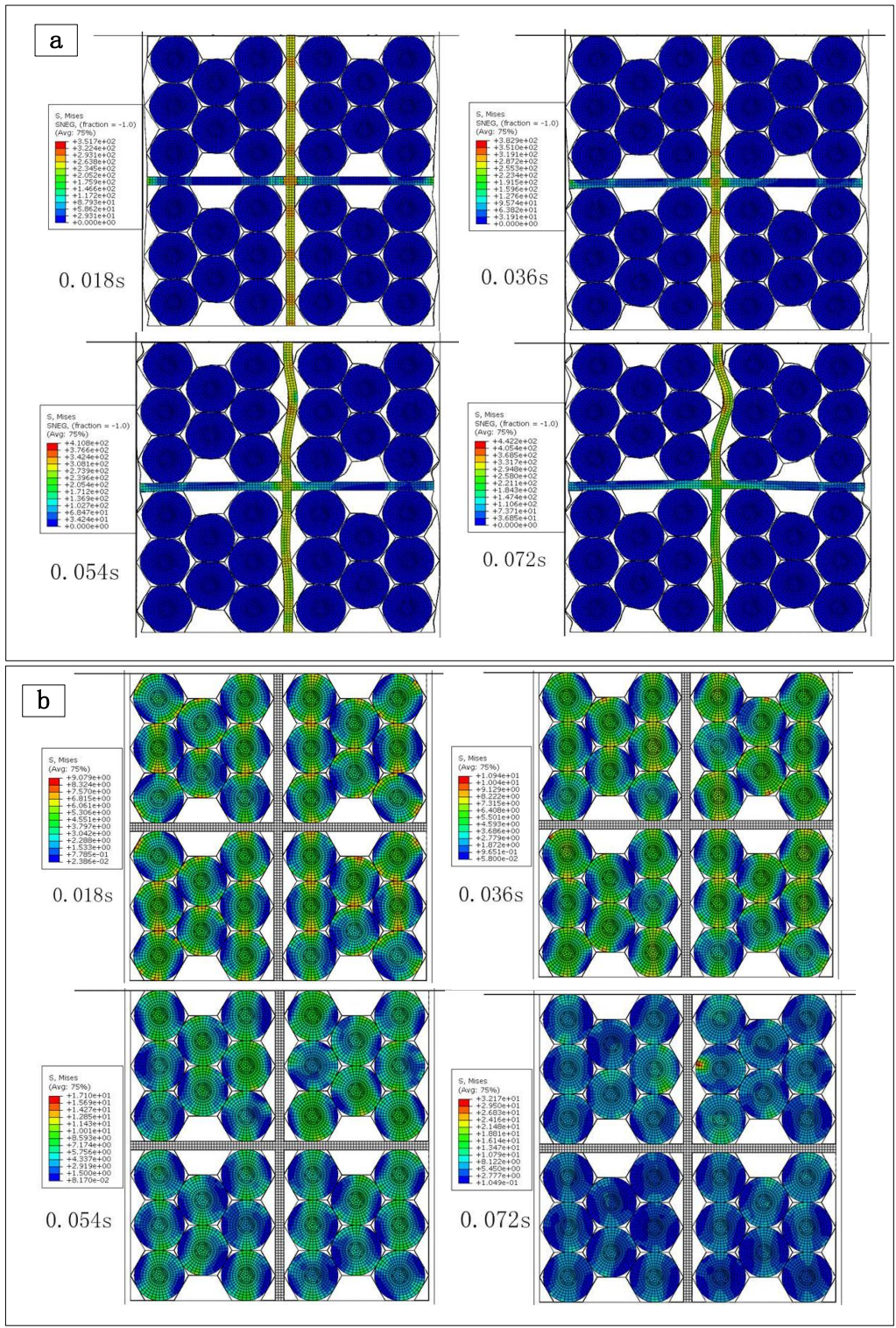

Fig. 9 The stress contours of fine FE model in different time (a) The stress contours of the whole package (b) The stress contours of the battery

## 3.3 Trust region space mapping

### 3.3.1 Introduction of TRSM algorithm

The TRSM is a mainstream SM algorithm. Its characteristic is setting a trust region for design parameter of fine model. The trust region of every cycle is used to judge whether the algorithm is convergent. Then, the TRSM will be introduced in detail.

Assuming $x$ is the design parameter of fine model and $f$ is the objective function of fine model. Similarly, the symbol $z$ and $c$ represent the design parameter and the objective function of coarse model, respectively. The optimization problem of the fine model is expressed as

$$\min \quad f(x) \tag{3}$$

The constraint function of the fine model is $g_1(x)$ which is expressed as

$$g^l \leq g_1(x) \leq g^u \tag{4}$$

$$x^l \leq x \leq x^u \tag{5}$$

where $g^l$ and $g^u$ are the lower and upper bounds of constraint, respectively, $x^l$ and $x^u$ are the lower and upper bounds of the design parameter $x$. The optimal solution of the fine model is expressed as

$$x^* = \mathop{\mathrm{argmin}}_{x \in \Omega^{(f)}} \quad f(x) \tag{6}$$

The optimal solution of the coarse model is expressed as

$$z^* = \mathop{\mathrm{argmin}}_{z \in \Omega^{(c)}} \quad c(z) \tag{7}$$

where $\Omega^{(f)}$ and $\Omega^{(c)}$ represent the design space of the fine model and the coarse model, respectively. The algorithm wants to establish the mapping relation between the two design spaces, that is $p: \Omega^{(f)} \rightarrow \Omega^{(c)}$. The establishment of the mapping relation is based on minimizing the residual. The expression is

$$p(x) = \mathop{\mathrm{argmin}}_{z \in \Omega^{(c)}} \ ||f(x) - c(z)|| \tag{8}$$

where $||f(x) - c(z)||$ is residual, and $||\cdot||$ is norm. According to the expression, if the residual is small enough, it follows

$$f(x) \approx c(p(x)) \tag{9}$$

If the parameters of two spaces satisfy

$$z^* = p(x^*) \tag{10}$$

a perfect mapping can be established. Therefore, the Eq. (6) can be transformed into

$$x^* \approx \mathop{\mathrm{argmin}}_{x \in \Omega^{(f)}} \ c(p(x)) \tag{11}$$

Now, the complex and expensive optimization problem is transformed into easy and cheap problem. The SM algorithm will solve Eq. (6) iteratively and solve Eq.(7) mainly at each iteration. At the $k$th cycles, the function $p(x)$ is replaced with linear approximation $p_k(x)$ which is expressed as

$$p_k(x) = B_k(x - x_k) + z_k \tag{12}$$

where $x \in \Omega^{(k)}$, $\Omega^{(k)} = \{x: ||x - x_k|| < \delta_k\}$. $B_k$ donates approximation Jacobin matrix of function $p_k(x)$. $\delta_k$ is the size of the optimization range in the $k$th cycles.

The specific steps of the algorithm are presented in Table 6. In every iteration, $x_{k+1}$ is accepted if it satisfies

$$\zeta = \frac{c(z_k) - c(z_{k+1})}{c(z_k) - c(p_k(x_{k+1}))} \tag{13}$$

The update rule of $B_k$ is

$$B_{k+1} = B_k + \frac{z_{k+1} - z_k - B_k h_k}{h_k^T h_k} h_k^T \tag{14}$$

where $h_k = x_{k+1} - x_k$.

The update rule of $\delta_k$ is

$$\delta_{k+1} = \begin{cases} 0.5 * \delta_k & \zeta \leq 0.01 \\ \delta_k & 0.01 \leq \zeta \leq 0.75 \\ \min(\delta_{max}, 2 * \delta_k) & \zeta \geq 0.75 \end{cases} \tag{15}$$

The algorithm will converge if it satisfied any one of these conditions:

$$\begin{cases} k \geq k_{max} \\ ||h_k|| \leq \varepsilon_1(1 + ||x_k||) \\ \delta_k \leq \varepsilon_1(1 + ||x_k||) \\ \left|\frac{f(x_{k+1}) - f(x_k)}{f(x_k)}\right| < \varepsilon_2 \\ c(z_k) - c(p_k(x))_{min} \leq 0 \end{cases} \tag{16}$$

where both $\varepsilon_1$ and $\varepsilon_2$ are constant coefficients, $k_{max}$ is the maximum number of cycles

Table 6 Descriptions of TRSM algorithm

| |
|---|
| Given $\delta_1$; set $B_1 = I(n,n), k = 1$; |
| $z^* = \underset{z \in \Omega^{(c)}}{\text{argmin}} \ c(z) \ ; x_1 = z^*;$ |
| Evaluate $f(x_1)$; |
| $z_1 = \underset{z}{\text{argmin}} \ ||f(x_1) - c(z)||$ ; Evaluate $c(z_1)$; |
| While 1: |

$$x_{k+1} = \underset{x \in \Omega^{(k)}}{\operatorname{argmin}} \ c(p_k(x)), \text{where } p_k(x) = B_k(x - x_k) + z_k, \Omega^{(k)} = \{x : ||x - x_k|| < \delta_k\} \ ;$$

Obtain $c(p_k(x))_{min}$

If algorithm converges:

    Break;

End;

$$z_{k+1} = \underset{z}{\operatorname{argmin}} \ ||f(x_{k+1}) - c(z)|| \ ;$$

Evaluate $c(z_{k+1})$;

Update $B_k$ and $\delta_k$;

Evaluate $f(x_{k+1})$;

If algorithm converges:

    Break;

End;

k = k + 1 ;

End;

### 3.3.2 Numerical example

In order to validate the superiority of TRSM algorithm, a simple function example is conducted. The fine model is expressed as

$$f(x) = \begin{cases} \cos(x) + 1 & \text{if } 2 \leq x \leq 9\pi/8 \\ -\cos(x - \pi + 2) \times 0.7 - 0.437 & \text{if } 9\pi/8 \leq x \leq 12 \end{cases} \quad (17)$$

The coarse model is expressed as

$$c(z) = -\left[\sin\left(2\pi \times \frac{z-1}{20} - 0.5\right)\right] \times 2.2 + 10 \quad z \in [2\ 12] \quad (18)$$

These two models are shown in Fig. 10. It is obvious that these two models are similar, but the coarse model is simpler. The objective function is

$$\min \ f(x) \quad (19)$$

There are two methods which are applied in the optimization problem. The one is TRSM algorithm whose optimization algorithm of coarse model is GA. The another one is GA which is used to optimize the fine model directly. When using TRSM algorithm to optimize the example, there are only one fine model and 230 coarse models are calculated. When the GA is used to optimize directly, the algorithm converges after 50 iterations. The result of two methods are shown in Table. 7. According to the result, the error between two optimal solutions which are obtained by TRSM and GA is small, and the computational cost of TRSM is much less than GA. Therefore, the TRSM algorithm is superior when the efficiency and accurate are taken into consideration together.

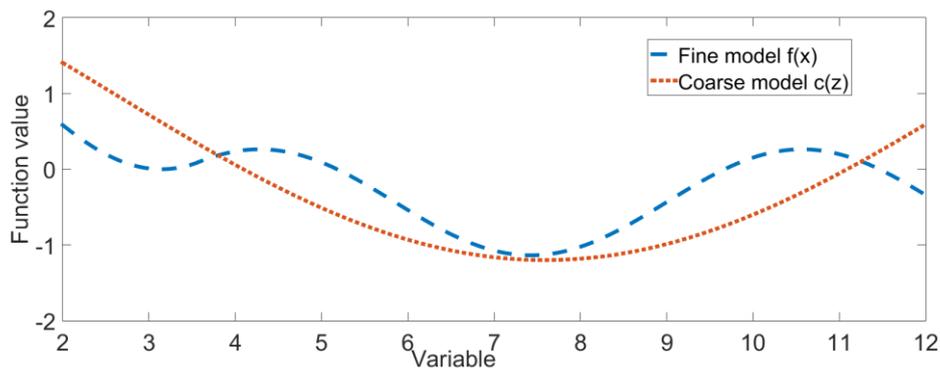

Fig. 10 The fine model and the coarse model

Table 7 The result of optimization of two algorithms

|  | $x^*$ | $f^*$ | Iterations | Time/s |
|---|---|---|---|---|
| TRSM algorithm | 7.5915 | -1.1273 | 1 | 0.039 |
| GA | 7.4248 | -1.137 | 50 | 2.23 |

## 3.4 The coarse FE model of battery package

In SM algorithm, an easier and cheaper FE model is needed which called coarse FE model. As shown in Fig. 8(c), the stress of battery is almost changeless in the axial direction of battery. Therefore, the FE model can be simplified as a pseudo-plane-strain model. The coarse FE model is shown in Fig. 8(d). All the translational degree of freedoms (DOFs) along the Z axis and the rotational DOFs along the X and Y axis are constrained. The simulation parameters of the coarse FE model are the same as the fine model. The whole model includes 10341 solid and 2080 shell elements. Figure. 11(a) shows the stress contours of coarse FE model in 0.018s, 0.036s, 0.054s and 0.072s, respectively. In order to show the integral stress contours of battery, the grid and honeycomb are excluded from the simulation output and only the battery is output as shown in Fig. 11(b). The maximum stress of battery is 15.1035MPa. The deformation of coarse FE model is similar to the deformation of fine FE model by comparing these two stress contours which in the same time. Because the coarse FE model needs only 5 minutes to finish a simulation when using the same CPUs and its stress states is similar to fine FE model, it can be an excellent coarse model of SM algorithms.

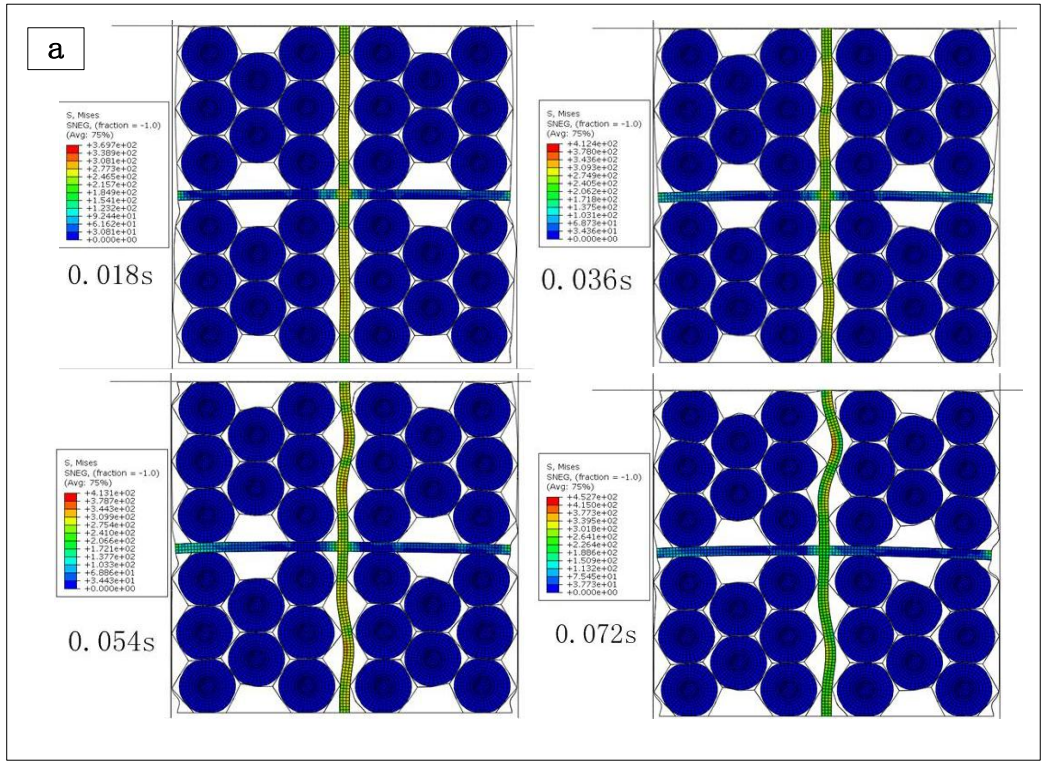

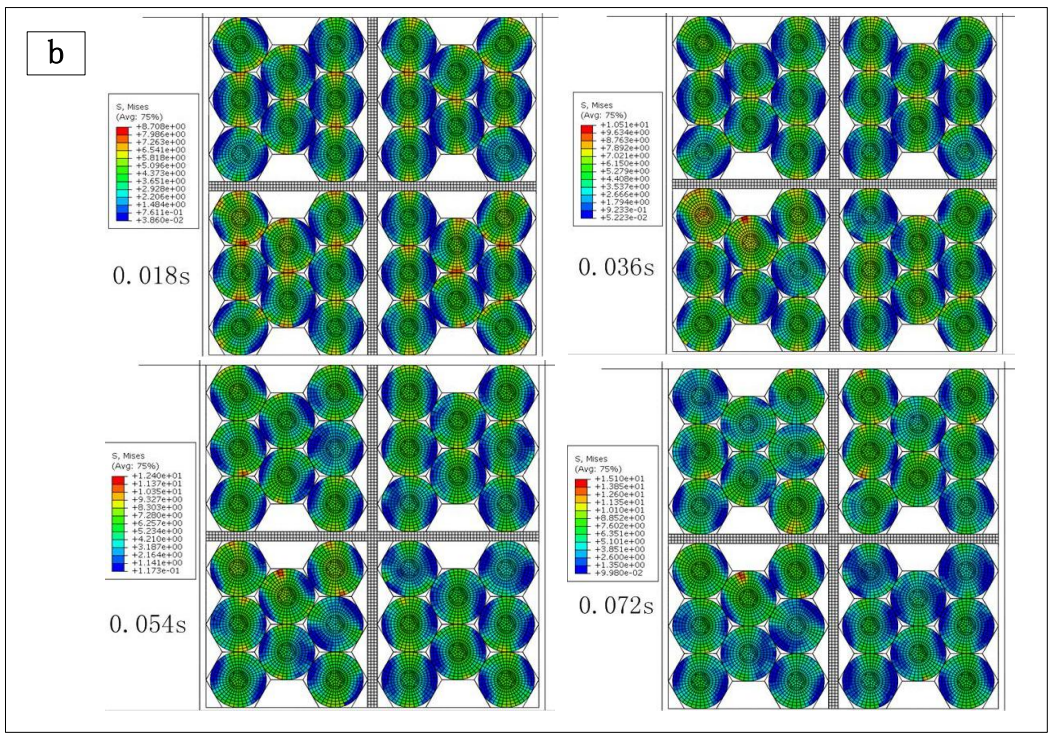

Fig. 11 (a) The stress contours of coarse FE model (b) The stress contours of the battery of coarse model in different time

## 3.5 Optimization of battery package

In the optimization problem, the design variables are the lengthways thickness of grid $h_1$, the lateral thickness of grid $h_2$ and the thickness of honeycomb core $h_3$. The range of these design variables are listed in Table 8. Because the stress of battery increases with time, the stress state of battery in the final time is concerned. In order to improve the magnitude and distribution of stress of batteries, the optimized target is suggested to minimize the mean and the variance of stress of all the batteries which in the final time. It is a multi-objective problem. The objective function is

$$\begin{cases} \min_{h1,h2,h3} \text{mean}(\sigma) \\ \min_{h1,h2,h3} \sigma^2(\sigma) \end{cases} \quad (20)$$

where $\sigma$ is the stress of all batteries in 0.072s. GA is used to optimize the coarse model. Ultimately, GA obtains a series of feasible solutions, and the solution with minimum square sum of mean and variance is selected be the optimum solution. Then, the optimum solution obtains from coarse model is used to build the mapping relation with fine model and then, we will obtain the optimum solution of fine model.

During the whole optimization, two fine models and 797 coarse models are calculated. The optimization costs 3 days to obtain a global optimal solution with 8 CPUS. With the same time, only 29 fine models could be calculated and it is impossible to obtain the global optimal solution. If the fine model was optimized by using GA, 230 fine models would be calculated at least and it would cost 24 days. Therefore, the TRSM algorithm is greatly helpful for the efficiency of this work.

The optimum solution of fine model is listed in Table 9. The design variable and objective history for the optimization problem are listed in Table 10. The norm of $c(p_k(x))$ of the first and the second cycle are shown in Fig. 12. According to the result listed in Table 10, the convergence condition is satisfied. What's more, due to the convergence judgment in the second cycle, the second cycle is interrupted. The algorithm obtains three $x$ ultimately. In the second cycle, the norm of $c(z_k)$ is less than the minimum of norm of $c(p_k(x))$. According to the Eq. (9), it can indicate the objective value is worse if $x = x_3$ and the globally optimal solution is not in the optimal range of the second cycle. Figure. 12 reveals that the result of the second cycle is much worse than the result of the first cycle. Therefore, the algorithm converges, and the optimal solution of first cycle is the globally optimal solution. The initial mean and variance of stress of fine model in 0.072s are 5.3231MPa and 5.9944, respectively. The optimized mean and optimized variance of stress of fine model are 3.5836MPa and 4.3904, respectively. Correspondently, The initial mean and variance of stress of coarse model in 0.072s are 5.2257MPa and 5.2641, and the optimized mean and variance of stress of coarse model are 3.6988MPa and 4.4957. Obviously, they close to the optimized solution of fine model. It means the fine model and the coarse model are similar essentially, therefore, the TRSM algorithm is greatly suitable for this work. Figure. 13 (a) shows the stress contours of the optimized fine model in 0.018s, 0.036s, 0.054s and 0.072s, respectively. Figure. 13(b) only shows the stress contour of coarse model in 0.072s. These stress contours only output the stress of battery. According to the figures, it is clear to know that the stress state of fine model and coarse model in 0.072s are similar. The maximum stress of batteries of fine model is 12.9642MPa in 0.072s. It drops by 59.71% compared to the maximum stress of initial fine model which is 32.1733MPa. The optimization consequence reveals that the magnitude and distribution of stress of batteries have been improved significantly.

Table 8 The range of design variables

| Parameters/mm | Range |
|---|---|
| The lengthways thickness of grid $h_1$ | [0.5,5] |
| The crosswise thickness of grid $h_2$ | [0.5,5] |
| The thickness of honeycomb core $h_3$ | [0.02,1] |

Table 9 The optimum solution of fine model

| Parameters/mm | Values |
|---|---|
| The lengthways thickness of grid $h_1$ | 2.0184 |
| The crosswise thickness of grid $h_2$ | 0.8023 |
| The thickness of honeycomb core $h_3$ | 0.03865 |

Table 10 The design variable and objective history for the optimization problem

| Cycles | Parameters $x_{k+1}$/mm | $c(z_{k+1})$ | $c(p_k(x))_{min}$ | $f(x_{k+1})$ |
|---|---|---|---|---|
| 0 | [1.6676 0.9112 0.0422] | [3.5765 4.6412] | / | [3.5263 4.6484] |
| 1 | [2.01840 0.80230 0.03865] | [3.5643 4.5399] | [3.5991,4.4420] | [3.5836 4.3904] |
| 2 | [1.4514 0.9041 0.0503] | / | [3.6112,4.5736] | / |

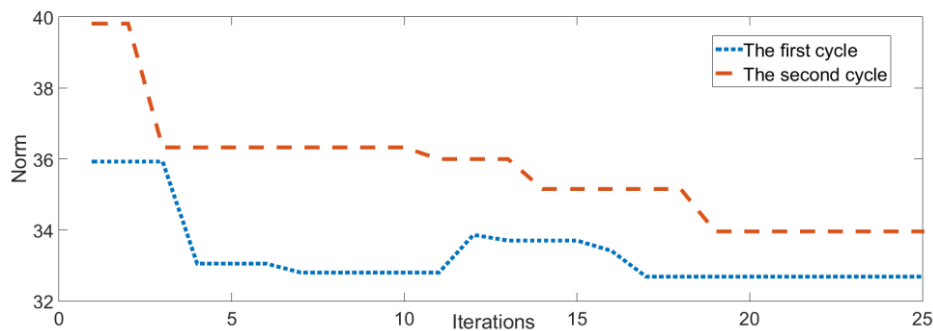

Fig. 12 The norm of $c(p_k(x))_{min}$ of the first and the second cycle

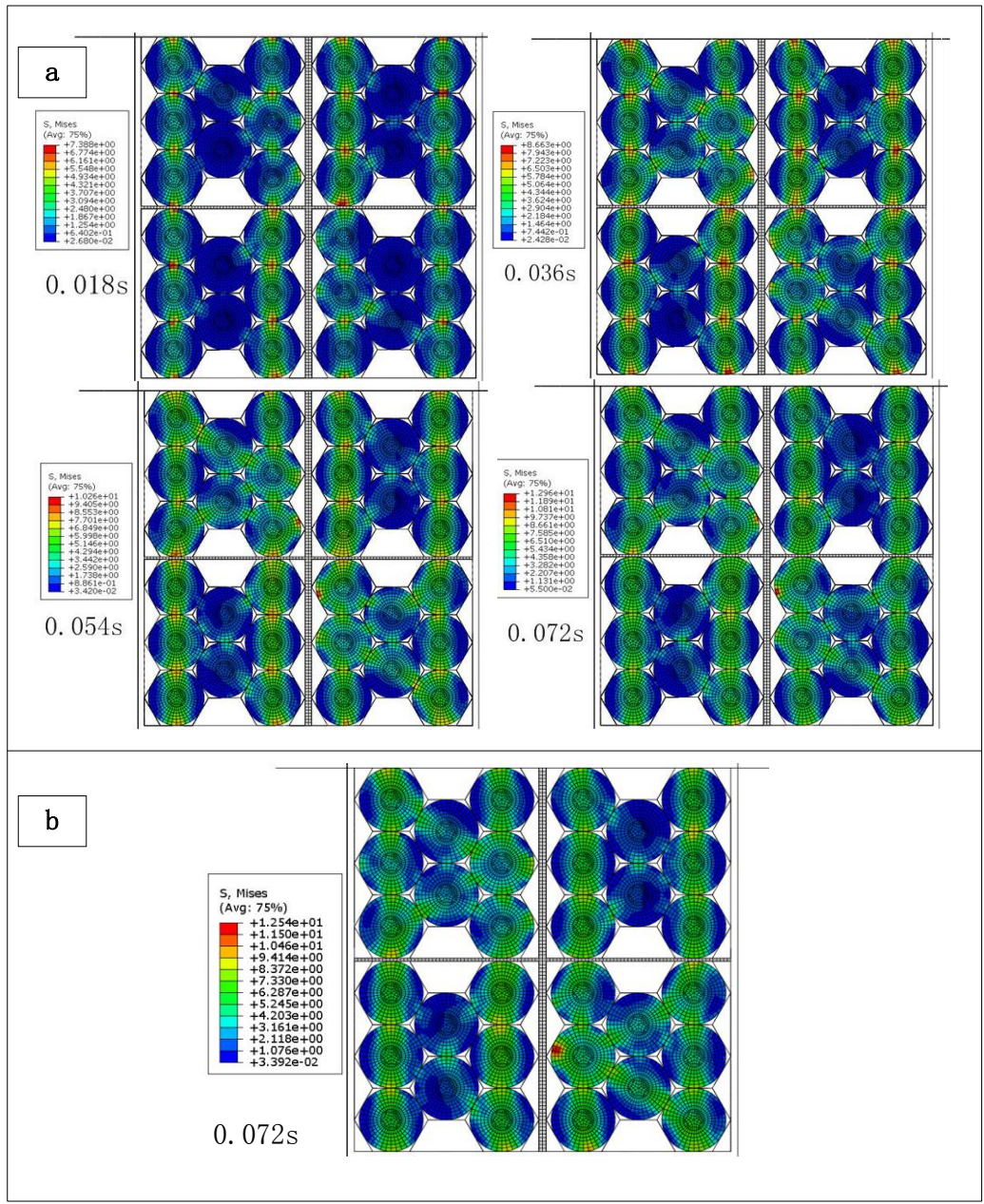

Fig. 13 (a) The stress contours of the optimized fine model (b) The stress contours of the optimized coarse model

## 4. Conclusions

A new honeycomb battery package structure is designed in this study. compared with poplar honeycomb structure, the proposed honeycomb structure uses a grid to reinforce its strength. In order to construct the FE model of the package, the material parameters of battery and Al alloy have to be obtained. Therefore, the flat compression test of 18650 battery and the uniaxial tensile test of Al alloy 6061-T6 are conducted. And then according to the data that obtained from flat compression test of battery, the material parameters of battery are identified by inverse identification method. Subsequently, the quasi-static compression test of battery package is conducted to validate the modeling method of battery package. The verification results reveal that the deformation shape of the test package sample and the deformation shape of FE model are similar. Therefore, the FE model of battery package is accurate. In order to obtain the optimized structure of package, TRSM algorithm is used to optimize the structure. Compared with other SM algorithms, the coarse model of SM is based on a pseudo-plane-strain model. Moreover, to guarantee the reliability, the mean and variance values of battery stress are used to be the objective function. Optimized consequence shows the maximum stress of inner battery is decreased from 32.1733MPa to 12.9642MPa, the mean of battery is decreased from 5.3231MPa to 3.5836MPa and the variance is decreased from 5.9944 to 4.3904. The optimization effect is quite remarkable. What's more, the computational cost of the optimization decreases significantly. Generally, the highlights of this work can be summarized as follows:

➢ A new structure of battery package is designed. It is a kind of honeycomb structure. In order to reinforce its strength, an Al alloy grid is used to be the framework.

➢ The material parameters of cylindrical battery are obtained by inverse identification method instead of replacing by material parameters of pouch battery.

- Due to the expensive cost of FE evaluation, the typical optimization method cannot be available for optimizing the battery package. Therefore, the TRSM algorithm can be used to optimize the structure and obtain an ideal result in an acceptable time. Compared with other application of the TRSM, the coarse model is specified modeled for the full FE package model.

## 5. Acknowledgments

This work has been supported by National Key R&D Program of China 2017YFB0203701, Project of the Key Program of National Natural Science Foundation of China under the Grant Numbers 11572120